\documentclass[printer]{aa} 
\usepackage{graphicx}
\usepackage{txfonts}
\usepackage{enumerate}
\usepackage{multirow}
\usepackage{lscape}
\usepackage{longtable}
\usepackage{color}
\def\fdeg{\hbox{$^\circ$}}
\let\footnote\thanks

\begin{document} 

\title{The 15273~\AA\ diffuse interstellar band in the dark cloud Barnard\,68 \footnote{Based on observations collected at the European Organisation for Astronomical Research in the Southern Hemisphere under ESO programme 096.C-0931(A)}}

\titlerunning{The 15723~\AA\ NIR-DIB in Barnard\,68}

\author{Meriem Elyajouri\inst{1}
        \and
        Nick L.J. Cox\inst{2} 
        \and
        Rosine Lallement\inst{1}
        }
\authorrunning{Elyajouri, Cox \& Lallement}

\institute{GEPI, Observatoire de Paris, PSL Research University, CNRS,
Place Jules Janssen, 92190 Meudon, France\\ \email{meriem.el-yajouri@obspm.fr}
\and
Anton Pannekoek Institute for Astronomy, University of Amsterdam, NL-1090 GE Amsterdam, The Netherlands}
\date{Received xx, 2017; accepted xx, 2017}
\abstract{High obscuration of background stars behind dark clouds precludes the detection of optical diffuse interstellar bands (DIBs) and hence our knowledge of DIB carriers in these environments. Taking advantage of the reduced obscuration of star-light in the near-infrared (NIR) we used one of the strongest NIR DIBs at 15273~\AA\ to probe the presence and properties of its carrier throughout the nearby interstellar dark cloud Barnard\,68. Equivalent widths (EW) have been measured for different ranges of visual extinction $A_V$, using VLT/KMOS H-band (1.46--1.85~$\mu$m) moderate-resolution (R$\sim$4000) spectra of 43 stars situated behind the cloud. To do so we fitted the data using synthetic stellar spectra from the APOGEE project and TAPAS synthetic telluric transmissions appropriate for the observing site and time period. 
The results show an increase of DIB EW with increasing $A_V$. However, the rate of increase is much flatter than expected from the EW-A$_{V}$ quasi-proportionality established for this DIB in the Galactic diffuse interstellar medium. Based on a simplified inversion assuming sphericity, it is found that the volume density of the DIB carrier is 2.7 and 7.9 times lower than this expected average value in the external and central regions of the cloud which have n$_H$ $\simeq\ 0.4$ and $3.5 \times\ 10^5$~cm$^{-3}$, respectively. Further measurements with multiplex NIR spectrographs should allow detailed modeling of such an \textit{edge effect} of this DIB and other bands and help clarifying its actual origin.}
\maketitle

\section{Introduction}
One of the longest standing spectroscopic mysteries in interstellar medium (ISM) studies is the identity of the carriers of the so-called diffuse interstellar bands (DIBs). Except for a convincing identification of C$_{60}^{+}$ \citep{Campbell2016}, DIB carriers are still unknown 
 and hypothesized to be due to long carbon chains, polycyclic aromatic hydrocarbons (PAHs) or dehydrogenated aromatic-rich moieties in the gaseous phase
\citep{herbig95,sarre2006,snow14, Jones2016}.
In contrast to the hundreds of optical (400--900~nm) DIBs \citep{Hobbs09} only about 20 to 30 near-infrared (NIR, 900--2500~nm) DIBs have been detected in the diffuse ISM (\citealt{Geballe11,2014A&A...569A.117C,Hamano16}).
These NIR DIBs offer an unique additional insight into the overall carrier population. In particular, in the NIR it is easier to probe more highly reddened lines-of-sight (i.e. $A_V > 10$~mag), either due to a longer column of dust in the line-of-sight to a distant background source or due to denser environments.
The absence of perfect correlations between different DIBs suggests that most bands arise from different carriers \citep{cami97}. Studies of DIBs in the optical reveal there is a clear relation between their presence and neutral and molecular hydrogen abundances as well as with the effective interstellar radiation field strength. The latter particularly affects relative band strength ratios \citep{vos11, Ensor17}.
\cite{Lan2015} have used massive SDSS datasets and demonstrated that most DIBs level off with respect to the reddening when sightlines cross opaque molecular clouds, and some extreme cases of DIB disappearance have also been found \citep{Snow2002}. This is also consistent with the weakness of NIR DIBs in dense interstellar environments (e.g. \citealt{1994MNRAS.268..705A}). Instead of looking at different sightlines that probe different interstellar clouds it is also possible to map the evolution of DIB carriers with extinction 
from the edge (representing diffuse ISM) to the core (representing dense ISM) of a \emph{single} dark cloud.\\
Tracing the behaviour of DIBs for $A_V$ up to 10$~mag$ is, at present, only practical when using strong DIBs in the NIR. The 15273~\AA\ DIB is particularly suited as its presence and properties in the ISM have been studied extensively thanks to the inclusion of this band in the APOGEE NIR survey (\citealt{Zasowski15,Elyajouri16,Elyajouri17}).\\
Here we present KMOS spectroscopic observations of the 15273~\AA\ DIB towards 43 background sources probing Barnard 68 (B68). B68 is a nearby (125~pc), relatively isolated, starless Bok globule conveniently situated against a backdrop of Galactic Bulge (giant) stars. It is often seen as the prototypical case of a Bonner-Ebert sphere and shows a radial averaged extinction law (spanning from a few to several tens of magnitudes in visual extinction) that follows very closely the predicted case (\citealt{2001Natur.409..159A}). The radial density and temperature profile of B68 have been reconstructed accurately through NIR extinction and far-infrared and sub-mm emission measurements (e.g. \citealt{2006ApJ...645..369B,Nielbock2012,Steinacker16}).
\section{Observations and data reduction}
\begin{figure}[t!]
  \centering
   \includegraphics[width=0.7\hsize]{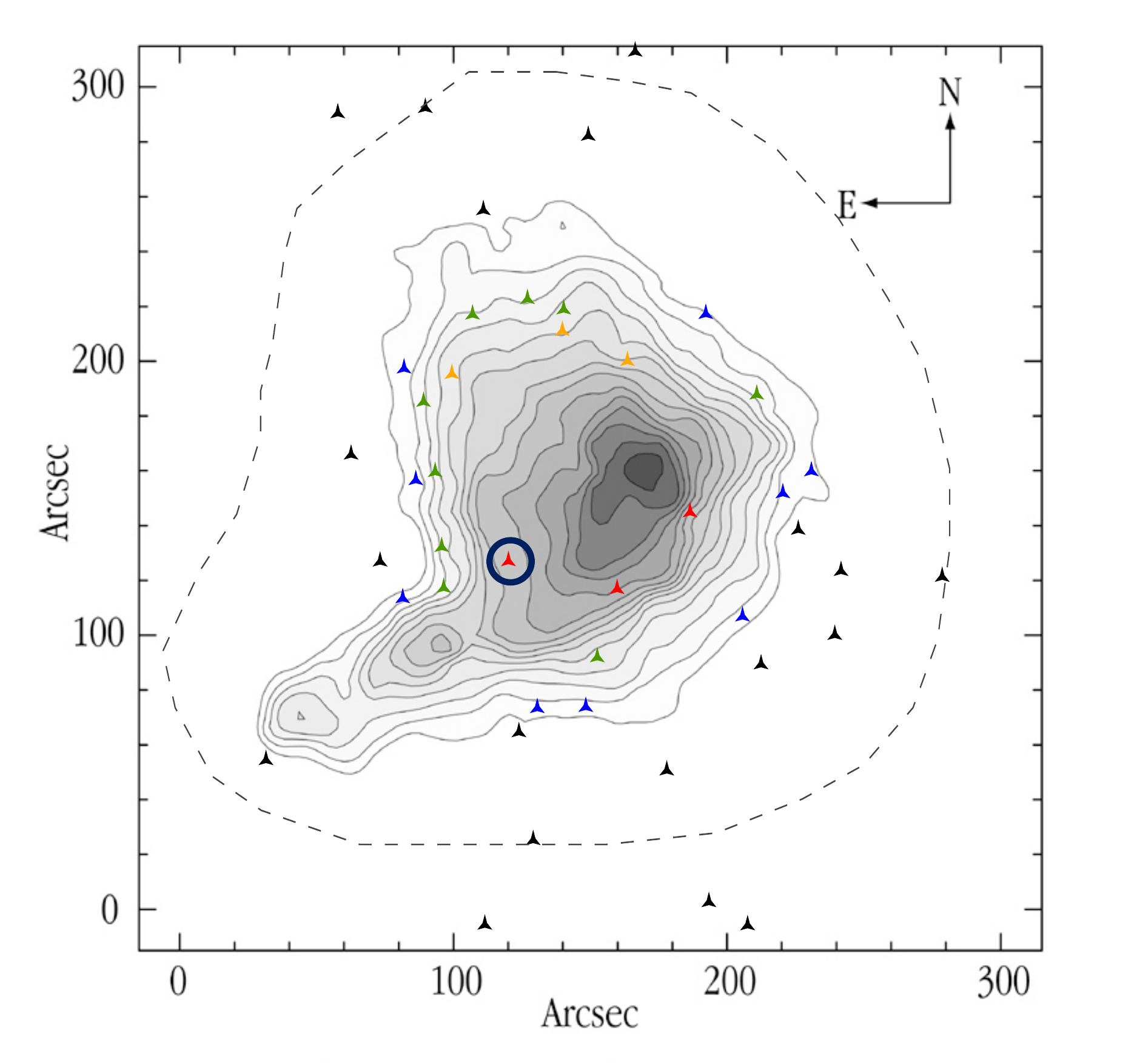}
      \caption{Target positions on the B68 extinction map from \citealt{2001Natur.409..159A}. Solid contours correspond to 2 mag steps from $A_V$= 4 to 30~mag. The dashed one is for $A_V$ =1.2~mag. Coordinates at center are 17h22m39s -23$\fdeg$50$\arcmin$00$\arcsec.$}
         \label{targets}
   \end{figure}

KMOS (\citealt{2013Msngr.151...21S}) $H$-band spectra of 85 targets towards B68 were obtained in February and March 2016 (Program ID 096.C-0931(A), PI: N. Cox).
The targets are located within $\sim$200\arcsec\ from the center of the cloud.
The 43 targets whose spectra could be used in the analysis (Sect.~\ref{fittingmethod}) are shown in Fig.~\ref{targets}.
The initial target list was based on selecting the 2MASS targets (J $\sim$ 12 -- 16~mag) at a range of distances from the center of the cloud. The final target selection was governed primarily by finding the most efficient configuration of the KMOS arms for the allocated time.
Four instrument set-ups were defined and executed.
Two configurations, in stare mode, were used to obtain spectra of the brightest target ($J$ = 12--14~mag) and two for the faintest targets ($J$ = 14--16 mag). 
The sky positions were observed every 3 science exposures. Parameters, seeing and airmass are listed in Table~\ref{tb:log}.
For each configuration about 20 arms were assigned to science targets and 3 to 4 arms were used for sky observations. The data were reduced using the KMOS data processing workflow \citep{Davies13} within the ESO Reflex environment. Average spectra were extracted from each 2.8\arcsec$\times$2.8\arcsec\ IFU.
In most cases optimal extraction was chosen. In two instances two sources were captured by the IFU and these were extracted separately using average extraction from specified spatial windows. All spectra are in the geocentric frame. The extracted spectra range from 1.425 to 1.867~$\mu$m (pixel size of 0.00021582~$\mu$m) and the spectral resolving power is R$\sim$4000 at 1.5273~$\mu$m. 
\begin{table}[h]
\caption{KMOS observations}\label{tb:log}
\centering                         
\begin{tabular}{lllll}        
\hline            
Setup                       & Seeing  & Airmass   & Date (2016) & Exp. time (s) \\
\hline                        
Bright \#1                  & 0.8\arcsec  & 1.57      & 02-29  &  160  \\
Bright \#2                  & 1.5\arcsec  & 1.17      & 02-21   &  160   \\
Faint \#1                    & 1.5\arcsec  & 1.53      & 03-17     &  420\\
Faint \#2                    & 2.7\arcsec  & 1.46      & 02-28      &  420\\
\hline  
\end{tabular}
\end{table}   

\section{Extraction of the interstellar spectrum \label{fittingmethod}}
\subsection{Synthetic stellar and telluric spectra}
In order to identify interstellar lines, one needs to fit each spectrum to a (stellar+telluric) model. Unfortunately, 2MASS stars behind B68 have unknown stellar parameters. To do a reliable analysis, synthetic models were extracted from the Apache Point Observatory Galactic Evolution Experiment (APOGEE; \cite{Majewski12}) Stellar Parameters and Chemical Abundances Pipeline (ASPCAP,
\citealt{2016AJ....151..144G}), which provides a synthetic stellar spectrum estimating the main stellar line locations for each giant star. In particular, we used models from the GK grid which covers 3500--6000~K \citep{meszaros12}.
From this set a sample of 1500 ASPCAP models, with $T_\mathrm{eff}$ = 3634 -- 6000~K, has been selected randomly. Since each APOGEE model is given in the corresponding stellar rest frame, the fitting procedure allows for a wavelength shift in order to compensate for the star radial velocity.
Telluric transmission models of H$_2$O adapted to the observing site, date and zenith angle were retrieved online from the Transmissions of the AtmosPhere for AStromomical data (TAPAS) service \citep{bertaux14}. 
The stellar and telluric spectra were convolved with a Gaussian instrumental response for R=4000. 
Fig.~\ref{b68005} shows an example of stellar spectrum and a typical telluric transmission  before and after convolution.
\subsection{Fitting method \label{sec:fit}}
\begin{figure}[t!]
    \includegraphics[width=\hsize]{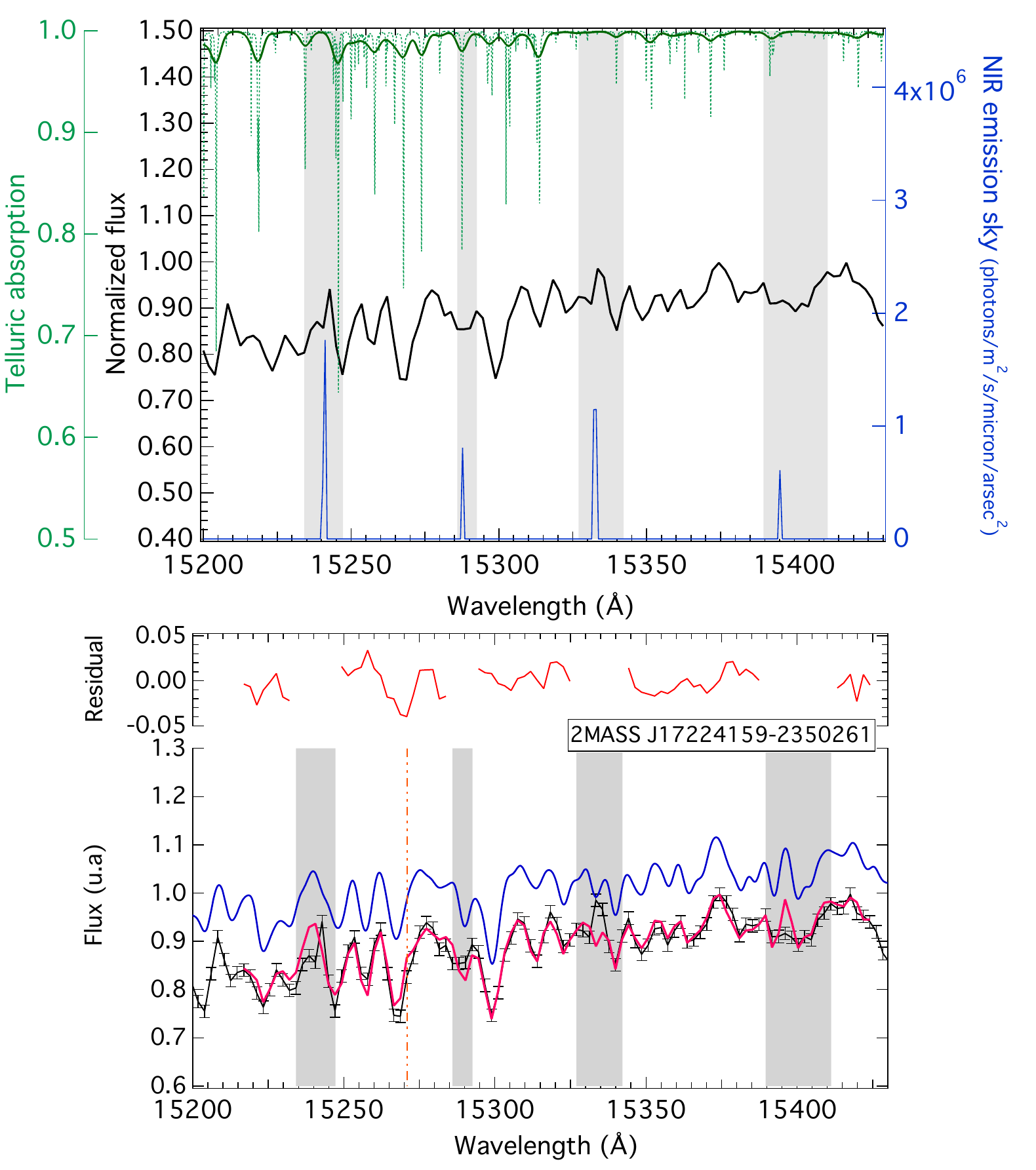}
      \caption{
      Top: Individual spectrum (black) for the circled target in Fig.~\ref{targets} and the green data point in Fig.~\ref{ewvsav}. The green lines show the telluric model before and after convolution. Gray rectangles indicate masks at strong sky emission lines (blue). Bottom: Example of fit. The blue curve is the convolved stellar model. The magenta curve is the fitted stellar+telluric model. The dashed orange line shows the DIB position. }
         \label{b68005}
   \end{figure}

\begin{figure}[t!]
\includegraphics[width=\hsize]{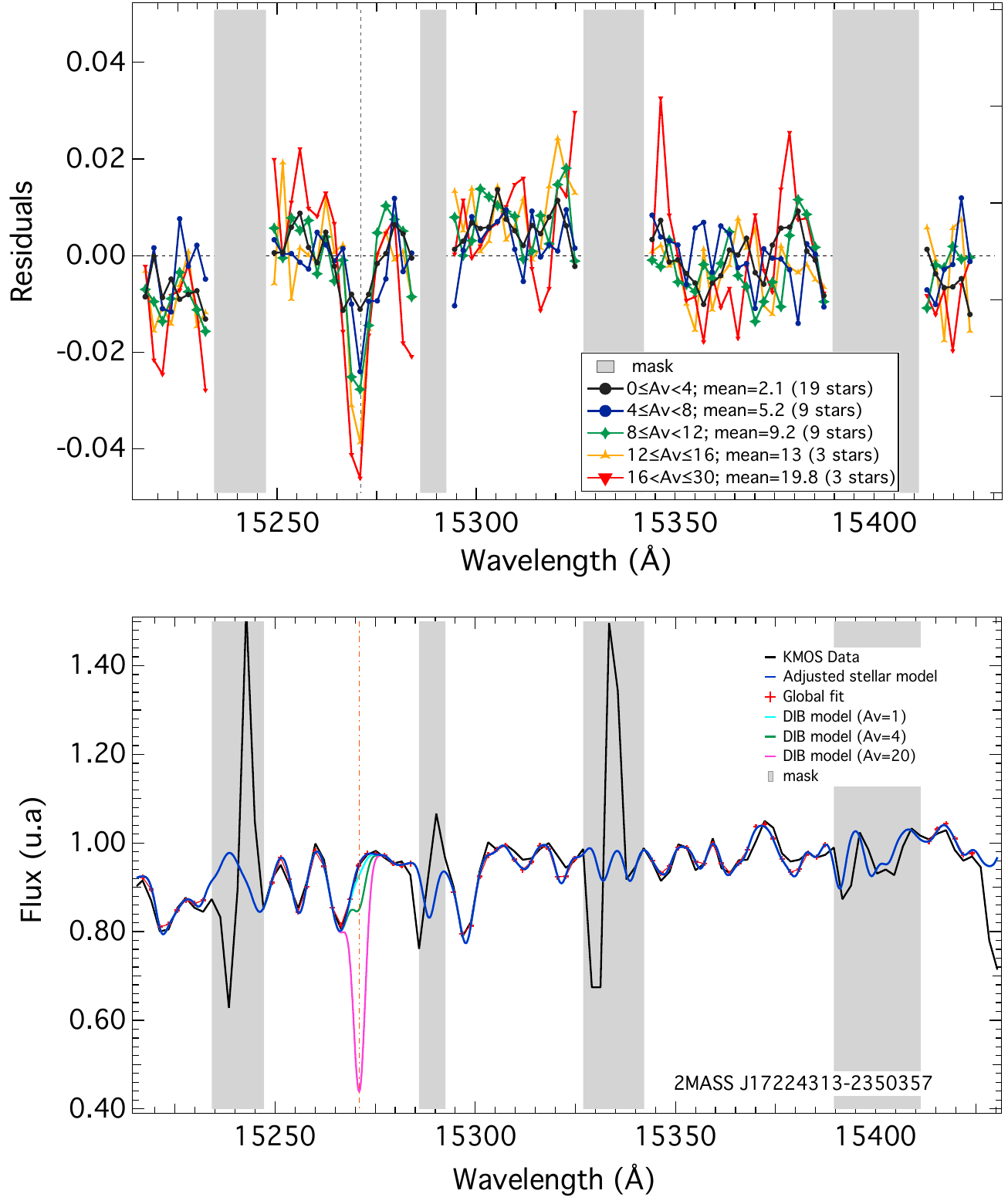}
 \caption{Top: residuals of model adjustments to the five average spectra (DIB excluded). Colors correspond to those in Fig.~\ref{targets}. Bottom: A target spectrum (in black) and its fitted model outside the DIB region (blue). Simulated Gaussian absorptions representing a DIB with the expected strength for $A_V$ = 1, 4 and 20 mag resp. are added.}
 \label{residu}
\end{figure}

Each individual spectrum has been fitted by a combination of the convolved synthetic stellar spectrum $S_\lambda$ and the convolved telluric transmission model T$_\lambda$ with a 8~\AA\ mask applied at the expected location of the DIB (around 15271~\AA). Note that "ill-fit" stellar lines regions have also been masked. A linear function (parametrized with $A$ and $B$) is included to adjust to the continuum level. Our model is described as follows :\\
$M_\lambda =
        (S_\lambda(\delta\lambda))^\alpha \times\ T_\lambda^\beta \times\ (A \times\ \lambda + B)$, 
 where $\alpha$ and $\beta$ are the scaling factors which adjust the stellar and telluric feature depths to the observed flux. $\delta \lambda$ is the free wavelength shift of the model. Note that the mathematically correct formula is the convolved product instead of our product of preliminarily convolved functions. However, we have checked that our simplified formula provides very similar fit results, owing to the broad instrumental function, but is much more efficient. As we focus on the strongest NIR DIB 15273, we selected a predefined fraction of the spectral range around the DIB location, i.e. 15216--15424~\AA. All parameters $A$, $B$, $\alpha$, $\beta$ and $\delta \lambda$ were free to vary. Note that the fit starts with the same initial guess except for the telluric scaling factor which is chosen according to the date of observation. We performed the adjustment for each of the 1500 ASPCAP models for each target and retained the model with the minimum $\chi^2$. 
We require the mean standard deviation to be less than 2.6\% outside the DIB, which yields a sample of 43 targets with clean and well-enough fitted spectra. A visual inspection of the residuals (R$_\lambda$ = observed flux - total best-fitting model)  allowed us to confirm the reliability of the procedure and that for selected targets the remaining fluctuations are solely due to the noise. These residuals were used to derive the actual noise level and error bars in individual spectra and in stacked spectra (see below). The SNRs in individual spectra range from 39 (i.e., our threshold) to $\simeq$125. Our method is illustrated in Fig.~\ref{b68005}. To check the feasibility of the DIB extraction we then performed a simulation using a single spectrum and its fitted model. DIB absorptions  were added with various strengths corresponding to the range of extinctions encountered in B68 and following the \cite{Zasowski15} relation  $\mathrm{EW} = 102\ \mathrm{m}\AA\ \times\ A_{V}^{(1.01 \pm 0.01)}$ derived over three orders of magnitude in extinction. Fig.~\ref{residu} shows that, in case the average law is verified, DIBs can be detected in this individual spectrum provided A$_{V}$ $\geq$~2. 
\section{15273~\AA~DIB characterization \label{DIBcharac}}
The residuals provide a first estimate of DIB detection or non-detection. However, we found that the DIB depth is often on the order of the noise amplitude. In order to improve the characterization of the DIB, we stacked residuals within groups of stars according to their A$_{V}$'s in five intervals centered on \{2.1; 5.2; 9.2; 13; 19.8\} as shown in Fig.~\ref{residu}. The corresponding numbers of targets in each group are $\{19; 9; 9; 3; 3\}$ and the SNRs decrease from $\simeq$170 to $\simeq$75.
Then, we performed two Gaussian fits at the DIB location. A Gaussian profile has been shown to be appropriate by\: \cite{Zasowski15} and \cite{Elyajouri17}. The first fit is done for a fixed DIB width and location, as if the DIB feature was due only to the dark cloud. In the second, both width and locations were free to vary. Uncertainties on EWs are estimated conservatively from the noise level as in \cite{Elyajouri16}. Fig.~\ref{ewvsav} displays the measured EW as a function of $A_V$ towards the B68 background star groups. Blue points show EWs with all free fit parameters and red ones correspond to the single cloud case. The results are compatible within error bars, but the fixed width case is found to correspond to a smoother evolution than the free width case, suggesting that this method is more reliable and from now we will adopt this solution. The results are compared with expectations from the above galactic average of \cite{Zasowski15}. It can be seen that the measured reddening values at cloud periphery and center would imply 2.7 and 6.8 times higher values.
\subsection{Simplified inversion}
\begin{figure}[t!]
 \includegraphics[width=0.8\hsize]{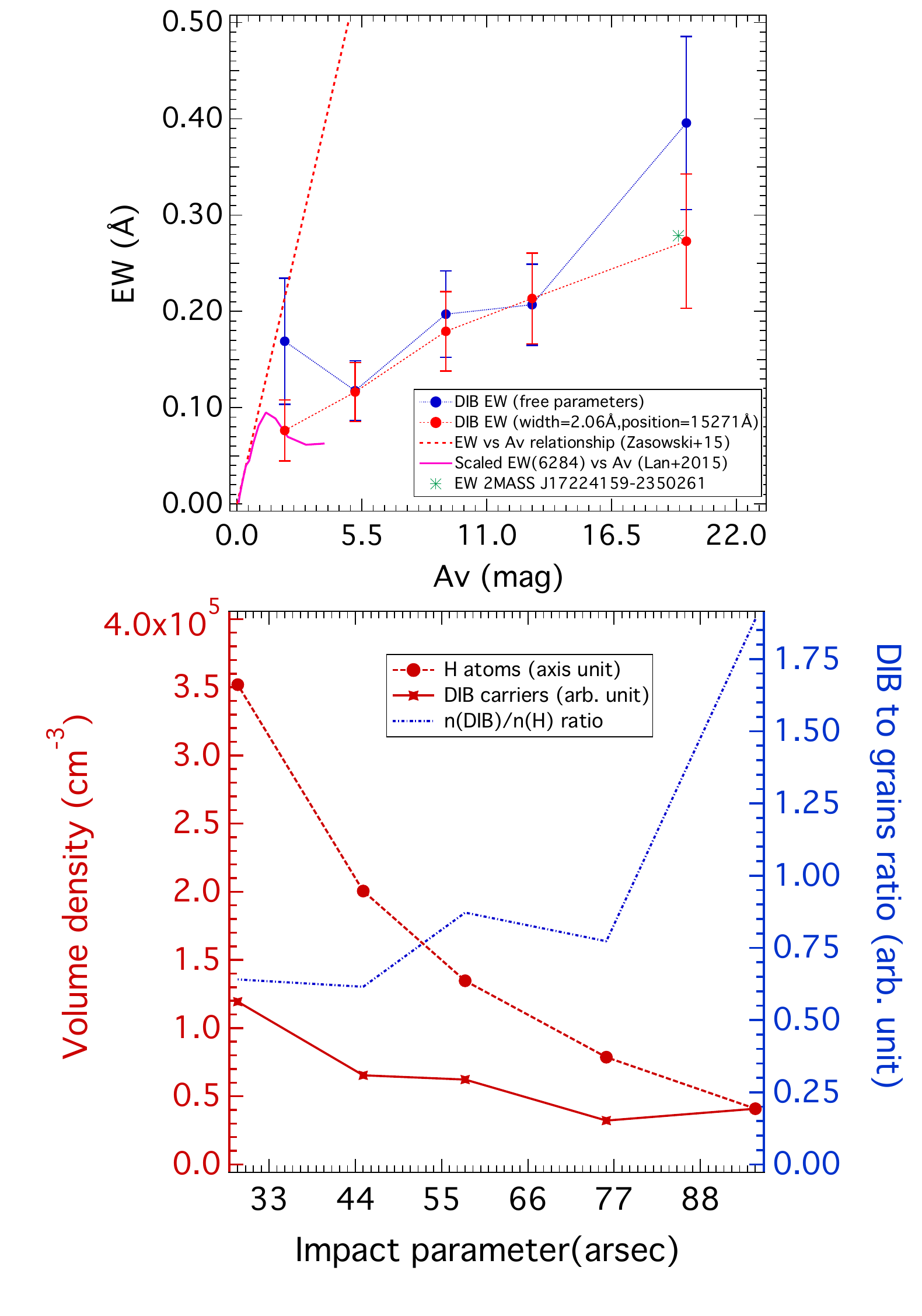}
\caption{Top: DIB EW versus $A_V$ for the five average spectra. 
Blue and red points correspond to the two fitting procedures (see text).
The dashed-red curve represents the \citet{Zasowski15} relation. The pink-line is the \citet{Lan2015} EW vs $A_V$ relationship for the 6284~\AA\ DIB, scaled to match the strength of the 15273~\AA~ DIB at $A_V\leq$1 (see text). Bottom: Volume density of DIB carrier n(DIB) based on the simplified inversion of EWs (left scale).  Also shown is the inverted volume density n(H) based on $A_V$ data. n(DIB) has been artificially scaled  in such a way the two curves cross for the external layer. The ratio between the DIB carrier and H volume densities is also displayed (right scale).}
 \label{ewvsav}
\end{figure}

We have performed a simplified inversion of DIB carrier columns (here EWs scaled by an unknown fixed number) in order to retrieve the true radial (3D) structure from the projected (2D) observed structure. We assumed that B68 is spherical and composed of concentric homogeneous spherical shells and the Abel transform can be replaced by an iterative computation starting from the most external layer. The density is assumed negligible outside of the most external layer. All parameters depend only on the radial distance r to the center that is proportional to the sightline angular distance from cloud center. The local volume density of the DIB carrier is described by a function n(DIB)(r) (cm$^{-3}$). The \textit{onion peeling} algorithm first computes n(r) in the external layer by simply dividing the external column density by the path length through the external shell. The next sightline through the adjacent, more internal shell, crosses the first and second shells only and the volume density in the second shell is obtained by dividing the DIB column by the path length through this second shell, after subtraction of the contribution of the path length through the first one, using the previously derived density. This simple computation is iterated until one reaches the cloud central densest sphere. Fig.~\ref{ewvsav} shows n(DIB) within the 5 shells (each corresponding to a measured EW). As a check of our procedure we have applied the same algorithm to the reddening, i.e. the column of large H nuclei assuming a constant dust to H ratio N$_{H}$= 6.25 $\times$ 10$^{25}$ E(B-V) and d=125 pc. The resulting volume densities n(H) range from 0.5 to 3.5 $\times$10$^{5}$ cm$^{-3}$ (Fig.~\ref{ewvsav}) and are roughly consistent with the results of \citet{Nielbock2012}, showing that despite the limited number of shells the procedure gives reasonable estimates of volume densities.
Fig.~\ref{ewvsav} also shows that the fractional abundance n(DIB)/n(H) decreases by a factor  $\simeq$~3 from periphery to center. A simple scaling shows that n(DIB)/n(H) is 2.7 (resp. 7.9) smaller than the average deduced from the \cite{Zasowski15} law at 0.018 (resp. 0.059) pc from cloud center.  
\section{Discussion}\label{secconclusion}
VLT KMOS spectra of stars beyond B68 have been used to derive the strengths of the associated 15273~\AA\ DIB, and to infer a first estimate of the volume density distribution of the DIB carrier. Following the general trend established for most optical DIBs (see e.g. \citealt{Lan2015}), the DIB EW is leveling off with respect to the extinction. At A$_{V}$=2.1 it is already at least 2.7 times below the average level measured along distant sightlines \citep{Zasowski15}, and, for our most central direction with A$_{V}\simeq$20, it is $\simeq$7 times below this level. In terms of volume densities in the spherical approximation, this corresponds to a 2.7 (resp. 7.9) depletion from periphery (r=0.018 pc) to center (r=0.059 pc). This 15273 ~\AA~ DIB depletion is less severe here than for strong optical DIBs in the remarkable case of the Orion star HD 62542 \citep{Snow2002}, in particular in the case of the 6284~\AA\ DIB (at least 25 times depleted with respect to average shielded $\zeta$-type sightlines, \cite{Adamk05}). This may be surprising since both DIBs have similar average responses to the radiation field \citep{Elyajouri17}. Fig.~ \ref{ewvsav} compares the dependence on the extinction of the 15273 ~\AA~ in B68 and of the 6284~\AA\ DIB in Fig. 9 of \cite{Lan2015}. A scaling factor has been applied to match the strength of DIB 15273~\AA\ in the linear region. There is a marked leveling off with respect to the extinction at A$_{V}\simeq$ 1 for both DIBs. Above A$_{V}$=2 the EW decrease observed for the 6284~\AA\ DIB may be absent for the 15273~\AA~ DIB, however the overlap in the extinction ranges is too small to firmly conclude. If confirmed, such differences would show that, despite the similarity between the responses to the radiation field, the 15273 DIB carriers do not behave in the dark cloud B68 as the 6284 DIB carriers in molecular clouds or Orion nebula stripped clouds. The life cycle of the carriers and its drivers are still under study. The UV radiation is recognized to play a primary role, and in the context of PAH-like carriers it would certainly directly influence the charge state of PAHs (e.g. \citealt{2005A&A...432..515R}) and the resulting disappearance of some DIBs. Additional photo-processing and hydrogenation or de-hydrogenation likely enter in play (e.g. \citealt{Vuong00}). \cite{BerLall17} suggested that a general carrier disappearance could additionally be due to coagulation into aggregates, as predicted by some evolutionary models of dust \citep{Jones2016}. Our study demonstrates that future measurements using multiplex NIR spectrographs and extended to more DIBs, more targets and various types of clouds should allow one to probe the volume density distribution of the DIB carriers precisely enough to shed light on the processes which govern their carrier formation and destruction.

\begin{acknowledgements}
We thank Jo\~ao Alves for providing us with the B68 extinction map and J.L. Bertaux for suggesting the \textit{onion peeling} algorithm.
M.E. acknowledges funding from the "Region Ile-de-France" through the DIM-ACAV project.
R.L. acknowledges support from "Agence Nationale de la Recherche" through the STILISM project (ANR-12-BS05-0016-02) and the CNRS PCMI national program.
\end{acknowledgements}
\bibliographystyle{aa} 
\bibliography{mybib}

\begin{thebibliography}{32}
\expandafter\ifx\csname natexlab\endcsname\relax\def\natexlab#1{#1}\fi

\bibitem[{{{\'A}d{\'a}mkovics} {et~al.}(2005){{\'A}d{\'a}mkovics}, {Blake}, \&
  {McCall}}]{Adamk05}
{{\'A}d{\'a}mkovics}, M., {Blake}, G.~A., \& {McCall}, B.~J. 2005, \apj, 625,
  857

\bibitem[{{Adamson} {et~al.}(1994){Adamson}, {Kerr}, {Whittet}, \&
  {Duley}}]{1994MNRAS.268..705A}
{Adamson}, A.~J., {Kerr}, T.~H., {Whittet}, D.~C.~B., \& {Duley}, W.~W. 1994,
  \mnras, 268, 705

\bibitem[{{Alves} {et~al.}(2001){Alves}, {Lada}, \&
  {Lada}}]{2001Natur.409..159A}
{Alves}, J.~F., {Lada}, C.~J., \& {Lada}, E.~A. 2001, \nat, 409, 159

\bibitem[{{Bergin} {et~al.}(2006){Bergin}, {Maret}, {van der Tak}, {Alves},
  {Carmody}, \& {Lada}}]{2006ApJ...645..369B}
{Bergin}, E.~A., {Maret}, S., {van der Tak}, F.~F.~S., {et~al.} 2006, \apj,
  645, 369

\bibitem[{{Bertaux} \& {Lallement}(2017)}]{BerLall17}
{Bertaux}, J.~L. \& {Lallement}, R. 2017, \mnras , in revision

\bibitem[{{Bertaux} {et~al.}(2014){Bertaux}, {Lallement}, {Ferron}, {Boonne},
  \& {Bodichon}}]{bertaux14}
{Bertaux}, J.~L., {Lallement}, R., {Ferron}, S., {Boonne}, C., \& {Bodichon},
  R. 2014, \aap, 564, A46

\bibitem[{{Cami} {et~al.}(1997){Cami}, {Sonnentrucker}, {Ehrenfreund}, \&
  {Foing}}]{cami97}
{Cami}, J., {Sonnentrucker}, P., {Ehrenfreund}, P., \& {Foing}, B.~H. 1997,
  \aap, 326, 822

\bibitem[{{Campbell} {et~al.}(2016){Campbell}, {Holz}, {Maier}, {Gerlich},
  {Walker}, \& {Bohlender}}]{Campbell2016}
{Campbell}, E.~K., {Holz}, M., {Maier}, J.~P., {et~al.} 2016, \apj, 822, 17

\bibitem[{{Cox} {et~al.}(2014){Cox}, {Cami}, {Kaper}, {Ehrenfreund}, {Foing},
  {Ochsendorf}, {van Hooff}, \& {Salama}}]{2014A&A...569A.117C}
{Cox}, N.~L.~J., {Cami}, J., {Kaper}, L., {et~al.} 2014, \aap, 569, A117

\bibitem[{{Davies} {et~al.}(2013){Davies}, {Agudo Berbel}, {Wiezorrek},
  {Cirasuolo}, {F{\"o}rster Schreiber}, {Jung}, {Muschielok}, {Ott}, {Ramsay},
  {Schlichter}, {Sharples}, \& {Wegner}}]{Davies13}
{Davies}, R.~I., {Agudo Berbel}, A., {Wiezorrek}, E., {et~al.} 2013, \aap, 558,
  A56

\bibitem[{{Elyajouri} {et~al.}(2017){Elyajouri}, {Lallement}, {Monreal-Ibero},
  {Capitanio}, \& {Cox}}]{Elyajouri17}
{Elyajouri}, M., {Lallement}, R., {Monreal-Ibero}, A., {Capitanio}, L., \&
  {Cox}, N.~L.~J. 2017, \aap, 600, A129

\bibitem[{{Elyajouri} {et~al.}(2016){Elyajouri}, {Monreal-Ibero}, {Remy}, \&
  {Lallement}}]{Elyajouri16}
{Elyajouri}, M., {Monreal-Ibero}, A., {Remy}, Q., \& {Lallement}, R. 2016,
  \apjs, 225, 19

\bibitem[{{Ensor} {et~al.}(2017){Ensor}, {Cami}, {Bhatt}, \& {Soddu}}]{Ensor17}
{Ensor}, T., {Cami}, J., {Bhatt}, N.~H., \& {Soddu}, A. 2017, \apj, 836, 162

\bibitem[{{Garc{\'{\i}}a P{\'e}rez} {et~al.}(2016){Garc{\'{\i}}a P{\'e}rez},
  {Allende Prieto}, {Holtzman}, {Shetrone}, {M{\'e}sz{\'a}ros}, {Bizyaev},
  {Carrera}, {Cunha}, {Garc{\'{\i}}a-Hern{\'a}ndez}, {Johnson}, {Majewski},
  {Nidever}, {Schiavon}, {Shane}, {Smith}, {Sobeck}, {Troup}, {Zamora},
  {Weinberg}, {Bovy}, {Eisenstein}, {Feuillet}, {Frinchaboy}, {Hayden},
  {Hearty}, {Nguyen}, {O'Connell}, {Pinsonneault}, {Wilson}, \&
  {Zasowski}}]{2016AJ....151..144G}
{Garc{\'{\i}}a P{\'e}rez}, A.~E., {Allende Prieto}, C., {Holtzman}, J.~A.,
  {et~al.} 2016, \aj, 151, 144

\bibitem[{{Geballe} {et~al.}(2011){Geballe}, {Najarro}, {Figer},
  {Schlegelmilch}, \& {de La Fuente}}]{Geballe11}
{Geballe}, T.~R., {Najarro}, F., {Figer}, D.~F., {Schlegelmilch}, B.~W., \& {de
  La Fuente}, D. 2011, Nature, 479, 200

\bibitem[{{Hamano} {et~al.}(2016){Hamano}, {Kobayashi}, {Kondo}, {Sameshima},
  {Nakanishi}, {Ikeda}, {Yasui}, {Mizumoto}, {Matsunaga}, {Fukue}, {Yamamoto},
  {Izumi}, {Mito}, {Nakaoka}, {Kawanishi}, {Kitano}, {Otsubo}, {Kinoshita}, \&
  {Kawakita}}]{Hamano16}
{Hamano}, S., {Kobayashi}, N., {Kondo}, S., {et~al.} 2016, \apj, 821, 42

\bibitem[{{Herbig}(1995)}]{herbig95}
{Herbig}, G.~H. 1995, ARA\&A, 33, 19

\bibitem[{{Hobbs} {et~al.}(2009){Hobbs}, {York}, {Thorburn}, {Snow}, {Bishof},
  {Friedman}, {McCall}, {Oka}, {Rachford}, {Sonnentrucker}, \&
  {Welty}}]{Hobbs09}
{Hobbs}, L.~M., {York}, D.~G., {Thorburn}, J.~A., {et~al.} 2009, ApJ, 705, 32

\bibitem[{{Jones}(2016)}]{Jones2016}
{Jones}, A.~P. 2016, Royal Society Open Science, 3, 160223

\bibitem[{{Lan} {et~al.}(2015){Lan}, {M{\'e}nard}, \& {Zhu}}]{Lan2015}
{Lan}, T.-W., {M{\'e}nard}, B., \& {Zhu}, G. 2015, \mnras, 452, 3629

\bibitem[{{Majewski}(2012)}]{Majewski12}
{Majewski}, S.~R. 2012, in American Astron. Soc. Meeting Abstracts 219, 205.06

\bibitem[{{M{\'e}sz{\'a}ros} {et~al.}(2012){M{\'e}sz{\'a}ros}, {Allende
  Prieto}, {Edvardsson}, {Castelli}, {Garc{\'{\i}}a P{\'e}rez}, {Gustafsson},
  {Majewski}, {Plez}, {Schiavon}, {Shetrone}, \& {de Vicente}}]{meszaros12}
{M{\'e}sz{\'a}ros}, S., {Allende Prieto}, C., {Edvardsson}, B., {et~al.} 2012,
  AJ, 144, 120

\bibitem[{{Nielbock} {et~al.}(2012){Nielbock}, {Launhardt}, {Steinacker},
  {Stutz}, {Balog}, {Beuther}, {Bouwman}, {Henning}, {Hily-Blant},
  {Kainulainen}, {Krause}, {Linz}, {Lippok}, {Ragan}, {Risacher}, \&
  {Schmiedeke}}]{Nielbock2012}
{Nielbock}, M., {Launhardt}, R., {Steinacker}, J., {et~al.} 2012, \aap, 547,
  A11

\bibitem[{{Ruiterkamp} {et~al.}(2005){Ruiterkamp}, {Cox}, {Spaans}, {Kaper},
  {Foing}, {Salama}, \& {Ehrenfreund}}]{2005A&A...432..515R}
{Ruiterkamp}, R., {Cox}, N.~L.~J., {Spaans}, M., {et~al.} 2005, \aap, 432, 515

\bibitem[{{Sarre}(2006)}]{sarre2006}
{Sarre}, P.~J. 2006, Journal of Molecular Spectroscopy, 238, 1

\bibitem[{{Sharples} {et~al.}(2013){Sharples}, {Bender}, {Agudo Berbel},
  {Bezawada}, {Castillo}, {Cirasuolo}, {Davidson}, {Davies}, {Dubbeldam},
  {Fairley}, {Finger}, {F{\"o}rster Schreiber}, {Gonte}, {Hess}, {Jung},
  {Lewis}, {Lizon}, {Muschielok}, {Pasquini}, {Pirard}, {Popovic}, {Ramsay},
  {Rees}, {Richter}, {Riquelme}, {Rodrigues}, {Saviane}, {Schlichter},
  {Schmidtobreick}, {Segovia}, {Smette}, {Szeifert}, {van Kesteren}, {Wegner},
  \& {Wiezorrek}}]{2013Msngr.151...21S}
{Sharples}, R., {Bender}, R., {Agudo Berbel}, A., {et~al.} 2013, The Messenger,
  151, 21

\bibitem[{{Snow}(2014)}]{snow14}
{Snow}, T.~P. 2014, in IAU Symposium, Vol. 297, The Diffuse Interstellar Bands,
  ed. J.~{Cami} \& N.~L.~J. {Cox}, 3--12

\bibitem[{{Snow} {et~al.}(2002){Snow}, {Welty}, {Thorburn}, {Hobbs}, {McCall},
  {Sonnentrucker}, \& {York}}]{Snow2002}
{Snow}, T.~P., {Welty}, D.~E., {Thorburn}, J., {et~al.} 2002, \apj, 573, 670

\bibitem[{{Steinacker} {et~al.}(2016){Steinacker}, {Bacmann}, {Henning}, \&
  {Heigl}}]{Steinacker16}
{Steinacker}, J., {Bacmann}, A., {Henning}, T., \& {Heigl}, S. 2016, \aap, 593,
  A6

\bibitem[{{Vos} {et~al.}(2011){Vos}, {Cox}, {Kaper}, {Spaans}, \&
  {Ehrenfreund}}]{vos11}
{Vos}, D.~A.~I., {Cox}, N.~L.~J., {Kaper}, L., {Spaans}, M., \& {Ehrenfreund},
  P. 2011, A\&A, 533, A129

\bibitem[{{Vuong} \& {Foing}(2000)}]{Vuong00}
{Vuong}, M.~H. \& {Foing}, B.~H. 2000, \aap, 363, L5

\bibitem[{{Zasowski} {et~al.}(2015){Zasowski}, {M{\'e}nard}, {Bizyaev},
  {Garc{\'{\i}}a-Hern{\'a}ndez}, {Garc{\'{\i}}a P{\'e}rez}, {Hayden},
  {Holtzman}, {Johnson}, {Kinemuchi}, {Majewski}, {Nidever}, {Shetrone}, \&
  {Wilson}}]{Zasowski15}
{Zasowski}, G., {M{\'e}nard}, B., {Bizyaev}, D., {et~al.} 2015, ApJ, 798, 35

\end{thebibliography}
\end{document}